\documentclass[epj]{svjour}
\usepackage{graphics}
\usepackage{graphics}
\usepackage{graphicx,float}
\usepackage[colorlinks,linkcolor=blue,citecolor=blue,urlcolor=blue,hyperindex]{hyperref}

\begin{document}
\title{Testing thermodynamic laws and weak cosmic censorship conjecture of conformal anomaly corrected AdS black hole}
\author{Guo-Ping Li\inst{1} % etc
% \thanks is optional - remove next line if not needed
\thanks{Corresponding author, e-mail: gpliphys@yeah.net} \and Ke-Jian He\inst{2} \and Bing-Bing Chen\inst{3}%
}                     % Do not remove
\offprints{}          % Insert a name or remove this line
\institute{College of Physics and Space Science, China West Normal University,
Nanchong, Sichuan 637002, People's Republic of China \and Department of Physics, Chongqing University, Chongqing 401331, People's Republic of China
\and Department of Physics, Si Chuan MinZu College, Kangding 626001, People¡¯s Republic of China}
\date{Received: date / Revised version: date}
% The correct dates will be entered by Springer
%
\abstract{
By dropping particles into black hole, we have employed the recently new assumption \cite{Page3} that the change of the black hole mass(enthalpy) should be the same amount as the energy of an infalling particle($\omega = dM$), to carefully test the laws of thermodynamics and the weak cosmic censorship conjecture of a conformal anomaly corrected AdS black hole in different phase spaces. Using the energy-momentum relation, the result shows that the first law, second law and weak cosmic censorship conjecture of black hole are all valid in the normal phase space, no violations occur. In the extended phase space, it firstly shows that the first law of black hole thermodynamics is always true in our case. Then, we interestingly find that if the condition ${d\ell} > -\left(P^r \ell^3\right)/{r_+^3}$ satisfied the variation of entropy is always positive, which means there would be no violation of the second law in the extend phase spaces. Also, it is true that there are always horizons by which the singularity is also covered. So, the configurations of the extremal and near-extremal black holes will not be changed, and there has no violation of the weak cosmic censorship conjecture. Finally, all of those conclusions are independent of the scalar curvature parameter $k$ and the conformal anomaly parameter $\tilde{\alpha}$.
 % end of PACS codes
} %end of abstract
\maketitle
\section{Introduction}
\label{sec1}
It is well known that
a black hole as a special object is defined by an event horizon, which is the boundary of the causal past of future null infinity. Classically, it indicates that the event horizon is an one-way channel with no energy or matter can reach an observer located outside of it. But in 1974, by considering the quantum effect, Hawking proved that a black hole can radiate particles, and possesses a temperature defined at event horizon \cite{Hawking,Hawking1}. It is because of this pioneering work, one realizes that the black hole can be regard as a thermodynamic system. Meanwhile, another thermodynamic quantity is found by Bekenstein, namely the black hole entropy which can be obtained from the irreducible mass, and it's value is proportional to the area of the horizon \cite{Bekenstein,Bekenstein1}. Combining the temperature and entropy, the laws of thermodynamics for black hole are subsequently established \cite{Hawking2}. Since then, the event horizon has been such an important concept that many researches with respect to the black hole cannot miss it \cite{1,2,3,4,6,7,8,93,94,92,9,91,10,11,12,13}. In particular, Penrose in 1969 proposed that the singularity of spacetime should be hidden by the event horizon of black hole in any real physical process and cannot be seen by distant observers, otherwise the causality of spacetime and all laws of physics will be broke down \cite{Penrose}. That is to say, there always exists a cosmic censorship which forbid the presence of a naked singularity. This proposal is the so-called ``weak cosmic censorship conjecture". Although the weak cosmic censorship conjecture is widely accepted for black holes, there is still lack of a general procedure to prove the validity of it.

In 1974, Wald put forward a Garden experiment to test the weak cosmic censorship conjecture \cite{Wald}. In the experiment, the aim of it is to check whether the event horizon of black hole is stable or not when a test particle with charge and angular momentum is captured by a black hole. He showed that a naked singularity is not allowed to appear for the extreme Kerr-Newman black hole since a test particle would not be captured by the black hole \cite{Wald}. Following this way, Hubeny reexamined the cosmic censorship of a near extremal Reissner-Nordstr\"{o}mblack hole, and found that the weak cosmic censorship conjecture is upheld \cite{Hubeny}. For a near-extremal Kerr black hole, Jacobson and Sotiriou presented that the black hole can be over-spun and the event horizon is destroyed by throwing a particle with the angular momentum into black hole \cite{Jacobson}. By further considering self-force effects and back reaction effects, Barausse and Isoyama have revised the weak cosmic censorship conjecture for a rotating black hole and a charged black hole, and found that there is no obvious evidence indicates the violation of the cosmic censorship in the proposed overcharging and
overspinning process \cite{Barausse,Barausse1,Isoyama}. In a word, more and more attention has been apaid to study this issue in the past decades and references therein \cite{Many1,Many2,Many3,Many4,Many5,Many6,Many7,Many8,Many9,Many10,Many11,Many14,Many12,Many13,Many16,Many17,Many15,Many18}.

By viewing $\Lambda$ as a dynamic variable, the pressure $P = -\Lambda / {8 \pi}$ is introduced as a thermodynamic quantity of black hole, and its conjugate quantity $V = ({\partial M} /{\partial P})_{S,Q,J}$ is found to be a thermodynamic volume \cite{Lammda1,Lammda2,Lammda3,Second,Lammda,Lammda4}. This idea was first considered by Henneaux and Teitelboim, and it has great physical significance to the AdS black hole thermodynamics \cite{Henneaux}. Firstly, the Smarr relation is consistent with the first law of thermodynamics in this extended phase space \cite{First}. Secondly, more fundamental theories which admit the variation of physical constants need to be further considered \cite{Second,First}. Thirdly, the mass of black hole is identified as enthalpy, rather than internal energy \cite{Third}. So in this extended phase space, one can see that the AdS black hole thermodynamics is naturally enriched, and then many interesting phenomena has been discussed, such as Van der Waals fluids, reentrant phase transitions, and holographic heat engines \cite{l1,l2,l3,l4,l5}.
In view of this, by considering the charged particle absorption, the study of the weak cosmic censorship conjecture in the extended phase space is a very interesting topic for AdS black hole.
In 2017, Gwak was the first to study the variation of the charged AdS black hole under charged particle absorption, and he found that the first law of thermodynamics and the weak cosmic censorship conjecture are still valid, but the second law of thermodynamics violated in the case considering thermodynamic volume \cite{Gwak}.
Then, when considering a fermion dropped into black hole, Chen has studied the thermodynamics and weak cosmic censorship conjecture in the extended phase spaces of AdS black holes in massive gravity \cite{Chen}, and his result is fully in consistence with that of \cite{Gwak}. Meanwhile, the thermodynamics and weak cosmic censorship conjecture with pressure and volume in the Gauss-Bonnet AdS black holes has been carefully addressed by Zeng, where the thermodynamic phase space is more extensive than that in previous studies \cite{Zeng}. It has been shown that the first law, second law as well as the weak cosmic censorship conjecture are all valid in the normal phase space, but the second law violated in the extended phase space when the first law and weak cosmic censorship conjecture are still valid \cite{Zeng}. Later on, this work has been extended to the case of the Kerr-AdS black hole, where the angular momentum $J$ are taken into account \cite{Zeng1}. In a word, the study of the black hole thermodynamics and the weak cosmic censorship conjecture with pressure and volume under particle absorption has attracted a lot of enthusiasm and has been investigated from various points of view in recent years \cite{Zeng2,Zeng3,Zeng4,Zeng5,Zeng6,Zeng7,Zeng8,Zeng9,Zeng10,Zeng11,Zeng12}.

Note that, it has been shown in \cite{Gwak,Chen,Zeng,Zeng1,Zeng2,Zeng3,Zeng4,Zeng5,Zeng6,Zeng8,Zeng7,Zeng9,Zeng11,Zeng12,Zeng10} that the second laws of black hole thermodynamics violated in the extended phase space. If this is true, one would be inevitably confronted with some unacceptable difficulties in physics: \textbf{(i)}, according to the AdS/CFT duality, this violation can give rise to a violation of the second law in the corresponding ordinary quantum field theory; \textbf{(ii)}, this violation would imply that there are something drastically wrong when one treats the cosmological constant as a variable. Moreover, it is generally believed that the laws that entropy always increases take up the supreme position among the laws of nature \cite{Natrue}. Therefore, Hu, Ong and Page have recently claimed that this putative violation of the second law of black hole in the extended phase space must be explained \cite{Page3}. And then, they have employed a new assumption that the energy of an infalling particle changes the mass(enthalpy) of the black hole by the same amount, to found that the first and second laws of a Reissner-Nordstr\"{o}m AdS black hole could be valid \cite{Page3}. However, we note in this case that, whether the weak cosmic censorship conjecture holds or not has not yet been carefully checked. Therefore, it is interesting to further study the thermodynamics and the weak cosmic censorship conjecture, thereby provide a piece of evidence for the assumption that the energy of an infalling particle indeed is the same amount as the change of the black hole mass(enthalpy).
On the other hand, the conformal anomaly is such an important role in quantum field theories that it has many applications which are closely related to black hole physics, cosmology, string theory, etc \cite{Cai}. And, many efforts have been devoted to study the effects of conformal anomaly on black hole thermodynamics \cite{Cai2,Cai1,Cai3,Cai4}. Although this effects on black hole thermodynamics have been widely discussed, it's unclear whether it has some interesting effects on the weak cosmic censorship conjecture in the extended phase space or not. Hence, it is necessary for us to further discuss the effects of conformal anomaly on the weak cosmic censorship conjecture with particle's absorption.
Motivated by those facts, our primary aim in this paper is to apply a new assumption to study the thermodynamics and weak cosmic censorship conjecture of the conformal anomaly corrected AdS black hole in two different phase spaces.
\\
\indent
The remainders of the present paper are outlined as follows. In Sec. \ref{sec2}, we review the thermodynamics of the conformal anomaly corrected AdS black hole, and then carefully investigate the relation between the energy and momentum of absorbed particles. Sec. \ref{sec3} is devoted to check the first law and second law as well as weak cosmic censorship conjecture in the normal phase. Sec. \ref{sec4} is aim to study the laws of the thermodynamics and the weak cosmic censorship conjecture in the extended phase space. Sec. \ref{sec5} ends up with a brief discussion and conclusion.

\section{\textbf{Motion of a charged particle in the conformal anomaly corrected AdS black hole}}\label{sec2}

In this section, we should first study the motion of a charged particle before testing the laws of thermodynamics and the weak cosmic censorship conjecture in the conformal anomaly corrected AdS black hole. In \cite{Cai2}, a newly derived conformal anomaly corrected AdS black holes has been obtained, which is\footnote{{Here, we have employed the units $G=\hbar=c=1$ throughout our paper.}}
\begin{equation}\label{q1}
ds^2 = -f(r)dt^2 + f(r)^{-1} dr^2 + r^2 d \Omega_{2 k}^2,
\end{equation}
where
\begin{equation}\label{q2}
f(r) = k - \frac{r^2}{4 \tilde{{\alpha}}} \left(1 - \sqrt{1 + \frac{8 \widetilde{\alpha}}{\ell^2} - \frac{16 \widetilde{\alpha} M}{r^3} + \frac{8 \widetilde{\alpha} Q^2 }{r^4}}\right).
\end{equation}
Here, the symbol $d \Omega_{2 k}^2$ represents the line element of a two-dimensional Einstein constant curvature space with scalar curvature, $\widetilde{\alpha}$ is a positive constant related to the content of the
conformal field theory, $A_t = Q/r_+$ is the non-vanishing component of the vector potential of this black hole, and two integration constants $M$ and $Q$ can be interpreted as the mass of black holes and the $U(1)$ conserved charge of the conformal field theory \cite{Cai2,Cai4}, respectively. The relation between the negative cosmological constant $\Lambda$ and the AdS radius $\ell$ is $\Lambda = -3 / {\ell^2}$, and the event horizon radius can be obtained by solving the equation $f(r_+) = 0$. Note that, the scalar curvature parameter $k$ is an important constant with it's value can be taken as 1, 0 or -1, which represents a positive, zero and negative constant curvature horizon of the conformal anomaly corrected AdS black hole \cite{Cai2,Cai4}.
In this paper, we will employ the Hamilton-Jacobi equation to find the energy-momentum relation of a charged particle near the event horizon. The Hamilton-Jacobi equation reads
\begin{equation}\label{q3}
\left(\partial^\mu S + q A^\mu \right)\left(\partial_\nu S + q A_\nu \right) + m^2 = 0,
\end{equation}
where $S$ is the action of particle, $A_\mu$ is the electromagnetic potential, $m$ and $q$ are the mass and charge of particle. By substituting the metric (\ref{q1}) into Eq.(\ref{q3}), we can get \footnote{For simplicity, the expression $d\Omega_{2k}$ has been replaced by $d\Omega_{2}$, where the reason is that $d\Omega_{2k}$ does not have any influence on the radial motion of a charged particle.}
\begin{eqnarray}\label{q4}
%\begin{aligned}
- \frac{1}{f(r)}\left(\partial_t S- q A_t \right)^2 + f(r)\left(\partial_r S  \right)^2 + \frac{1}{r^2}\left(\partial_\theta S  \right)^2 +\frac{1}{r^2 \sin{\theta}^2} \left(\partial_\phi S  \right)^2 + m^2 = 0,
%\end{aligned}
\end{eqnarray}
For the symmetries of the spacetime, carrying on the separation of variables as $ S = - \omega t + W(r) + \Theta(\theta) + L \phi$ for the conformal anomaly corrected AdS black hole, we have
\begin{equation}\label{q5}
\omega + q A_t  =  \sqrt{\left(P^r \right)^2 + f(r)\left(\frac{K}{r^2}   + m^2 \right)},
\end{equation}
where $P^r \equiv f(r)P_r(r) = f(r) \partial_r W(r)$ is the radial momentum of the particle, $\omega$ and $L$ are the energy and angular momentum of the particle, respectively. And, the equation (\ref{q5}) is the relation between the momentum, energy and charge of the ingoing particle, in which the angular part defined as $K = \left(\partial_\theta \Theta \right)^2 + \frac{ 1}{{\sin{\theta}}^2 L^2}$. To check the laws of thermodynamics and weak cosmic censorship conjecture, the energy-momentum relation (\ref{q5}) near the event horizon can be simplified as
\begin{equation}\label{q6}
\omega + q A_t  = | P^r |.
\end{equation}
In Eq.(\ref{q6}), it is noteworthy that one should choose the positive sign in front of the $| P^r |$ term to ensure the positive flow of time direction of a particle when it fell into the black hole \cite{Gwak}.

\section{\textbf{Thermodynamics and weak cosmic censorship conjecture in the normal phase space}}\label{sec3}

\begin{figure}[htp]
\begin{minipage}[t]{0.5\linewidth}
  \centering
  \setlength{\abovecaptionskip}{-0.1 cm}
  \setlength{\belowcaptionskip}{-0.4 cm}
  %\begin{minipage}[c]{0.5\textwidth}
  %\centering
  \includegraphics[width=0.65\textwidth,height=4.4cm]{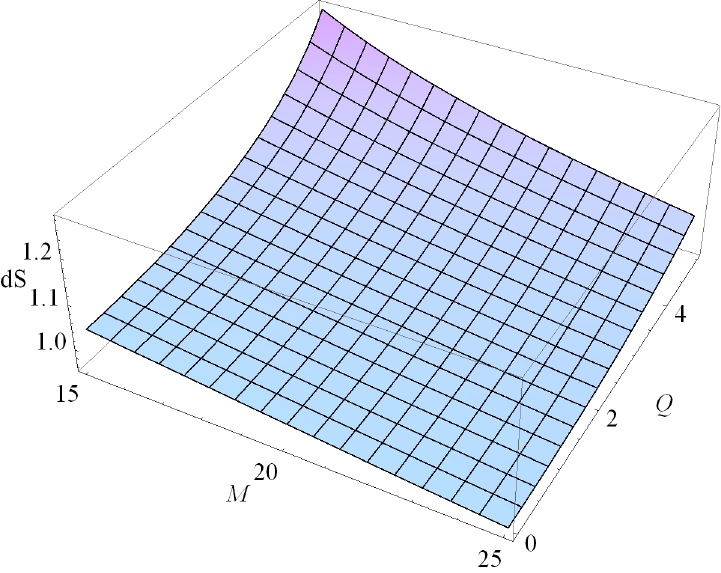}\\
  \caption{\scriptsize{The relation among $dS$, $M$ and $Q$ for the case $k=1, \widetilde{\alpha}=0.5$.}}\label{fig1}
 %\end{minipage}
 %\caption{²¢ÅÅͼÐÎ}
\end{minipage}%
\begin{minipage}[t]{0.5\linewidth}
  \centering
  \setlength{\abovecaptionskip}{-0.1 cm}
  \setlength{\belowcaptionskip}{-0.4 cm}
  %\begin{minipage}[c]{0.5\textwidth}
  %\centering
  \includegraphics[width=0.65\textwidth,height=4.4cm]{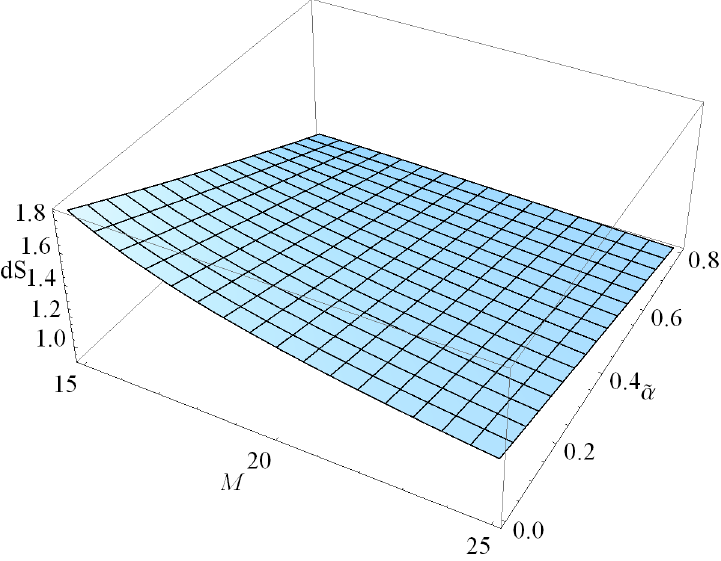}\\
  \caption{\scriptsize{The relation among $dS$,$M$ and $\widetilde{\alpha}$ for the case $k=1, Q=5$.}}\label{fig2}
 %\end{minipage}
 %\caption{²¢ÅÅͼÐÎ}
 \end{minipage}%
\end{figure}

\begin{figure}[htp]
\begin{minipage}[t]{0.5\linewidth}
  \centering
  \setlength{\abovecaptionskip}{-0.1 cm}
  \setlength{\belowcaptionskip}{-0.4 cm}
  %\begin{minipage}[c]{0.5\textwidth}
  %\centering
  \includegraphics[width=0.65\textwidth,height=4.4cm]{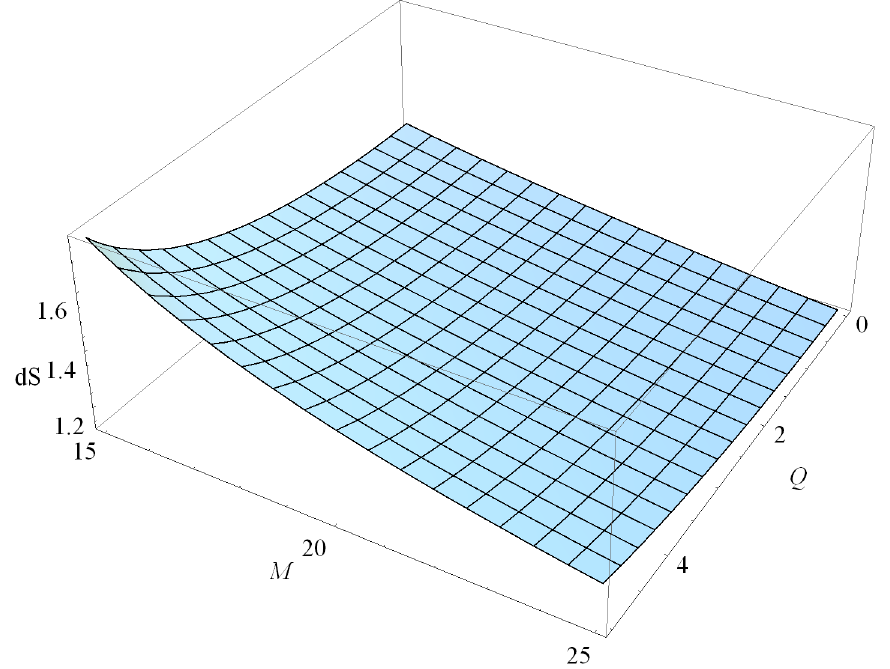}\\
  \caption{\scriptsize{The relation among $dS$, $M$ and $Q$ for the case $k=0$.}}\label{fig4}
 %\end{minipage}
 %\caption{²¢ÅÅͼÐÎ}
\end{minipage}%
\begin{minipage}[t]{0.5\linewidth}
  \centering
  \setlength{\abovecaptionskip}{-0.1 cm}
  \setlength{\belowcaptionskip}{-0.4 cm}
  %\begin{minipage}[c]{0.5\textwidth}
  %\centering
  \includegraphics[width=0.65\textwidth,height=4.4cm]{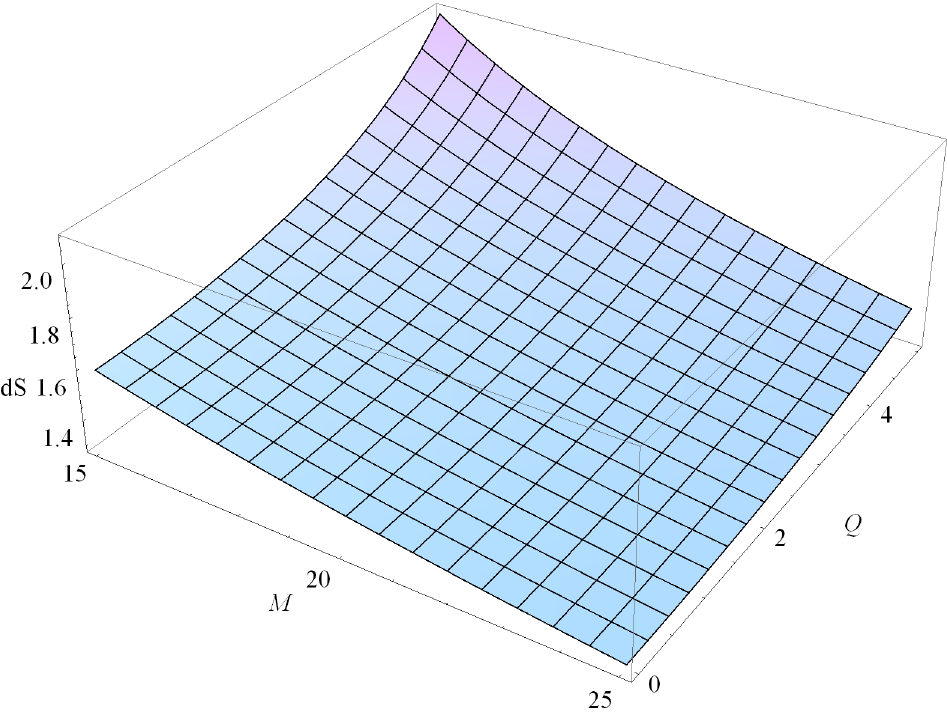}\\
  \caption{\scriptsize{The relation among $dS$, $M$ and $Q$ for the case $k=-1, \widetilde{\alpha}=0.5$.}}\label{fig5}
 %\end{minipage}
 %\caption{²¢ÅÅͼÐÎ}
\end{minipage}%
\end{figure}

\begin{figure}[htp]
  \centering
  \setlength{\abovecaptionskip}{-0.1 cm}
  \setlength{\belowcaptionskip}{-0.4 cm}
  %\begin{minipage}[c]{0.5\textwidth}
  %\centering
  \includegraphics[width=0.35\textwidth,height=4.4cm]{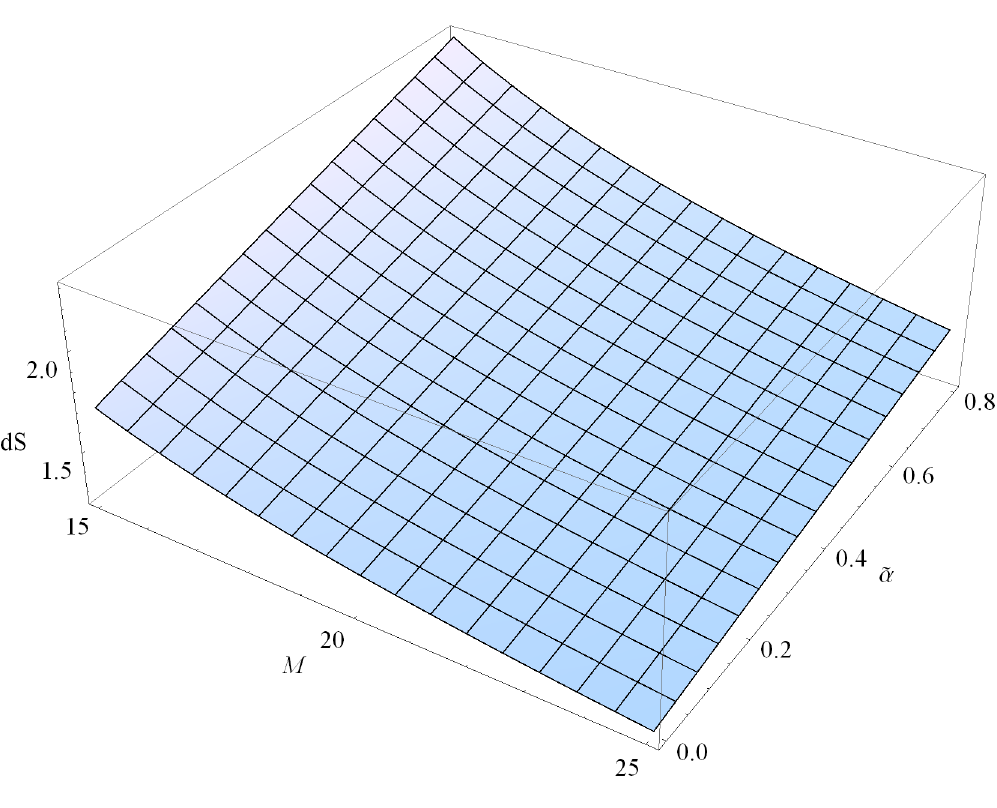}\\
  \caption{\scriptsize{The relation among $dS$, $M$ and $\widetilde{\alpha}$ for the case $k=-1, Q=5$.}}\label{fig6}
 %\end{minipage}
 %\caption{²¢ÅÅͼÐÎ}
\end{figure}

In this section, we will employ the energy-momentum relation (\ref{q6}) to check the laws of thermodynamics and weak cosmic censorship conjecture in the normal phase space by considering the particle's absorption. The electrostatic potential difference between the black hole horizon and the infinity can be expressed as,
\begin{equation}\label{q7}
\Phi  = \frac{Q}{r_+}.
\end{equation}
Based on the event horizon radius $r_+$, the expression of the mass of this black hole is
\begin{equation}\label{q8}
M = \frac{r_+^4 + k \ell^2 r_+^2 - 2 k^2 \ell^2 \widetilde{\alpha} +  \ell^2 Q^2}{2 \ell^2 r_+}.
\end{equation}
For the conformal anomaly corrected AdS black hole, the Hawking temperature of this black hole reads
\begin{equation}\label{q9}
T = \frac{f'(r_+)}{4 \pi} = \frac{r_+}{4 \pi (r_+^2 - 4 k \widetilde{\alpha})} \left( k + \frac{3 r_+^2}{\ell^2} - \frac{Q^2}{r_+^2} + \frac{2 k^2 \widetilde{\alpha}}{r_+^2}  \right),
\end{equation}
and the entropy is obtained as
\begin{equation}\label{q10}
S = \int \frac{1}{T}\left( \frac{\partial M}{\partial r_+} \right) dr_+ = \pi r_+^2 - 8 \pi k  \widetilde{\alpha} \ln r_+.
\end{equation}
when a charged particle dropped into the black hole, the change of black hole parameter should be exactly equal to that of the infalling particle, if one assuming no loss of conserved quantities during this process \cite{Gwak}. In this sense, the relation of the internal energy and charge between the black hole and the particle has the form as,
\begin{equation}\label{q11}
 \omega = dM, q = dQ .
\end{equation}
In this case, Eq.(\ref{q6}) change into
\begin{equation}\label{q12}
dM = \Phi dQ + P^r .
\end{equation}
As a charged particle dropped into the black hole, the configurations of the black hole will be changed. This progress will lead to a shift for the horizon, namely $dr_+$. Near the new horizon, the relation $f(r_+ + dr_+) = 0 $ is also satisfied. This means, $f( dr_+) = 0 $, which form is given by
\begin{equation}\label{q13}
df_+ = \frac{\partial f_+}{\partial M} dM + \frac{\partial f_+}{\partial Q} dQ + \frac{\partial f_+}{\partial r_+}dr_+ =0,
\end{equation}
where $df_+ = f( dr_+)$. By substituting Eq.(\ref{q12}) into Eq.(\ref{q13}), it's easy to found that the terms $dM$, $dQ$ are all eliminated. Therefore, we can obtain the value of the $dr_+$, which is
\begin{equation}\label{q14}
dr_+ = \frac{P^r r_+^2}{-2Q^2 + 3Mr_+ - k r_+^2 + 4 k^2 \widetilde{\alpha}} .
\end{equation}
With the aid of Eq.(\ref{q10}), the variation of entropy is given by
\begin{equation}\label{q15}
dS = \frac{2 P^r \pi r_+ (r_+^2 - 4 k \widetilde{\alpha})}{-2Q^2 + 3Mr_+ - k r_+^2 + 4 k^2 \widetilde{\alpha}} .
\end{equation}
Combining Eq.(\ref{q9}) with Eq.(\ref{q15}), there is a relation expressed as
\begin{equation}\label{q16}
T dS =  P^r .
\end{equation}
Inserting this relation into Eq.(\ref{q12}), then,
\begin{equation}\label{q17}
dM = \Phi dQ + TdS .
\end{equation}
Obviously, the expression (\ref{q17}) is exactly the first law of the black hole thermodynamics. It can be seen that the first law of the black hole thermodynamics also holds true in the normal phase space when the black hole captured a particle.
In the next, we will check the second law of the black hole thermodynamics by particle's absorption. The second law states that the entropy of the black hole never decrease in the clockwise direction. Based on this notion, one can check the second law of the black hole thermodynamics by studying the variation of entropy $dS$. For the extremal black hole where it's temperature is zero. Then, combining this condition and the black hole mass (\ref{q8}), the variation of entropy (\ref{q15}) finally reads,
\begin{equation}\label{q18}
dS_{extremal} \rightarrow\infty ,
\end{equation}
where we have used the condition $r_+^2 - 4 k \widetilde{\alpha} > 0$. So, it is true from Eq.(\ref{q18}) that the second law of black hole thermodynamics is still hold for the extremal black holes. For the non-extremal black holes, it's temperature is larger than zero. This indicates\footnote{For the case $k=0$, there is no existence of the condition $r_+ > 2 \sqrt{k \widetilde{\alpha}}$. }
\begin{equation}\label{q19}
  r_+ > 2 \sqrt{k \widetilde{\alpha}}, \quad Q^2 < r_+^2 k + 2 k^2 \widetilde{\alpha} + \frac{3 r_+^4}{ \ell^2}.
\end{equation}
After carrying the condition (\ref{q19}) on Eq.(\ref{q15}), it is certain that the variation of entropy $dS$ always has a positive value, which means the second law of black hole thermodynamics dose not violate for the non-extremal black holes. To clearly show $dS$, we plot Figs.\ref{fig1}-\ref{fig6} to present the value of $dS$ for the non-extremal black holes, where some acceptable parameters are fixed, i.e., $\ell=1, p^r=1$.


For a black hole with a positive curvature horizon, i.e., $k=1$, it shows from Figs.\ref{fig1},\ref{fig2} that the value of $dS$ increases with the black hole charge $Q$, but decreases with the black hole mass $M$ and the conformal anomaly parameter $\widetilde{\alpha}$. For a black hole with a Ricci
flat horizon, i.e., $k=0$, one can easily check that the conformal anomaly parameter $\widetilde{\alpha}$ vanished in Eq.(\ref{q15}), thereby it has no influence on $dS$. Meanwhile, we in Fig.\ref{fig4} find that the variation of entropy in this case is also bigger and bigger with the black hole charge $Q$, but smaller and smaller with the black hole mass $M$. For a black hole with a negative constant curvature horizon, i.e., $k=-1$, it is obvious from Figs.\ref{fig5},\ref{fig6} that the change of $dS$ with the mass and charge $(M,Q)$ is consistent to  that found in the case $k=1$. However, one can see in Fig.\ref{fig6} that the variation of entropy indeed increase with the parameter $\widetilde{\alpha}$, which is contrary to that found in the case $k=1$. Although the value of $dS$ could be influenced by those parameters, the fact that it is a positive value has never changed. In the next, we will continue to study the weak cosmic censorship conjecture in a conformal anomaly corrected AdS black hole with particle's absorption.

The weak cosmic censorship conjecture states that the singularity of spacetime should be hidden by the event horizon of black hole in any real physical process. To test it, an effective method is to check whether the event horizon of black hole is stable or not when a charged particle is captured by a black hole. However, the black hole horizon is determined by $f(r)$. Therefore, testing the weak cosmic censorship conjecture is to check whether the equation $f(r) = 0$ has solutions or not by considering particle's absorption. For the conformal anomaly corrected AdS black hole, assuming that it is at the radial coordinate $r_m$ that the matric function $f(r)$ has a minimum value $f(r_m)$. So, if $ f(r_m) > 0 $, there has no horizon for black hole; if $ f(r_m) \leq 0 $, there are horizon always. At the location $r_m$, the condition for holding the weak cosmic censorship conjecture is
\begin{eqnarray}\label{q21}
& f(r)|_{r = r_m}  \equiv f_m = \delta \leq 0 ,\\
& \partial_r f(r)|_{r = r_m}  \equiv f'_m = 0, \\
& (\partial_r)^2 f(r)|_{r = r_m}  \equiv f''_m > 0.
\end{eqnarray}
For the extremal black hole, we have $ \delta = 0, r_+ = r_m$. For the near extremal black hole, $\delta$ is a small and negative quantity, and $r_m$ is located between the inner horizon and outer horizon. When a charged particle fall into the black hole, the mass $M$ and charge $Q$ of black hole will be replaced by $M +dM, Q + dQ$, respectively. Naturally, the value of $r_m$ change into $r_m + dr_m$ and the change of the function $f(r_m)$ is
\begin{eqnarray}\label{q22}
df_m = f(r_m + dr_m) - f_m = \frac{\partial f_m}{\partial M}dM + \frac{\partial f_m}{\partial Q}dQ,
\end{eqnarray}
where the condition $f'_m= 0 $ has been employed in Eq.(\ref{q22}). For the extremal black hole, the horizon $r_+$ is equal to the radial coordinate $r_m$, so the relation (\ref{q12}) can be used to calculate the shift $df_m$. Inserting Eq.(\ref{q12}) into Eq.(\ref{q22}), the shift $df_m$ can be finally expressed as the following form with the aid of the expression of $(M,Q)$, which is
\begin{equation}\label{q23}
df_m = - \frac{2 P^r r_m}{r_m^2 - 4 k \widetilde{\alpha}},
\end{equation}
where $r_m^2 - 2 k \widetilde{\alpha} > 0$, and the temperature of the extremal black hole is equal to zero. So, combining Eq.(\ref{q16}) with Eq.(\ref{q23}), we have
\begin{equation}\label{q24}
f_m + df_m = 0.
\end{equation}
From Eq.(\ref{q24}), it is true that the value of $f_m + df_m$ is equal to zero. That is to say, the black hole has a horizon when it captured a charged particle, which means the configurations of the extremal conformal anomaly corrected AdS black hole have not been changed.
\\
\indent
For the near-extremal black hole, the key relation (\ref{q12}) is no longer true at the coordinate $r_m$ because it only holds true at $r_+$. Two locations $(r_+, r_m)$ are very close for the near-extremal black holes. So, we assuming the condition $r_+ = r_m + \epsilon,$ thereby $\delta$ becomes the minimum value $\delta_\epsilon$, where $\epsilon$ and $\delta_\epsilon$ are all the very small quantity. Using this condition, the relation (\ref{q12}) can be expand at the location $r_m$, it yields
\begin{eqnarray}\label{q25}
dM = &&k dr_m + \frac{\left(  M dr_m - QdQ  \right) \epsilon}{r_m^2}  + \frac{ QdQ - Mdr_m}{r_m}  \\
&&+ \frac{4 r_m \epsilon  dr_m }{\ell^2}  +\frac{ 2 r_m^2 dr_m}{\ell^2}  + O(\epsilon)^2,
\end{eqnarray}
Substituting Eq.(\ref{q25}) into Eq.(\ref{q22}), it leads to
\begin{equation}\label{q26}
df_m = O(\epsilon)^2.
\end{equation}
So, at the new minimum point, we have
\begin{equation}\label{qq26}
f_m + df_m =\delta_\epsilon +  O(\epsilon)^2.
\end{equation}
From the equation (\ref{q26}), it is obvious that there are the two horizons for the near-extremal black hole. And, the value of $f_m+df_m$ for the extremal black hole can be reproduced by setting $\epsilon =0, \delta_{\epsilon} = 0$ in (\ref{qq26}). So, in the normal phase space, one can see that the weak cosmic censorship conjecture also holds true for the near-extremal black hole, which is full in consistence with that obtained in \cite{Gwak,Chen,Zeng,Zeng1,Zeng2,Zeng3,Zeng4,Zeng5}.

\section{\textbf{Thermodynamics and weak cosmic censorship conjecture in the extended phase space}}\label{sec4}

\begin{figure}[htbp]
\begin{minipage}[t]{0.5\linewidth}
  \centering
  \setlength{\abovecaptionskip}{-0.1 cm}
  \setlength{\belowcaptionskip}{-0.4 cm}
  %\begin{minipage}[c]{0.5\textwidth}
  %\centering
  \includegraphics[width=0.65\textwidth,height=4.4cm]{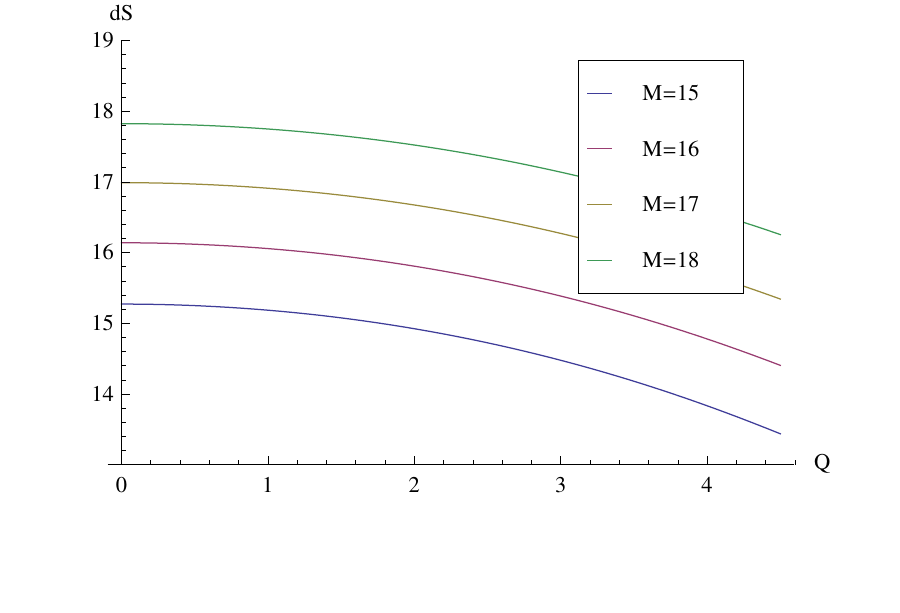}\\
  \caption{\scriptsize{The relation among $dS$, $M$ and $Q$ for the case $k=1,  \protect\\
  \widetilde{\alpha} =0.5, \ell=1, d\ell=0.5$.}}\label{f1}
 %\end{minipage}
 %\caption{²¢ÅÅͼÐÎ}
\end{minipage}%
\begin{minipage}[t]{0.5\linewidth}
  \centering
  \setlength{\abovecaptionskip}{-0.1 cm}
  \setlength{\belowcaptionskip}{-0.4 cm}
  %\begin{minipage}[c]{0.5\textwidth}
  %\centering
  \includegraphics[width=0.65\textwidth,height=4.4cm]{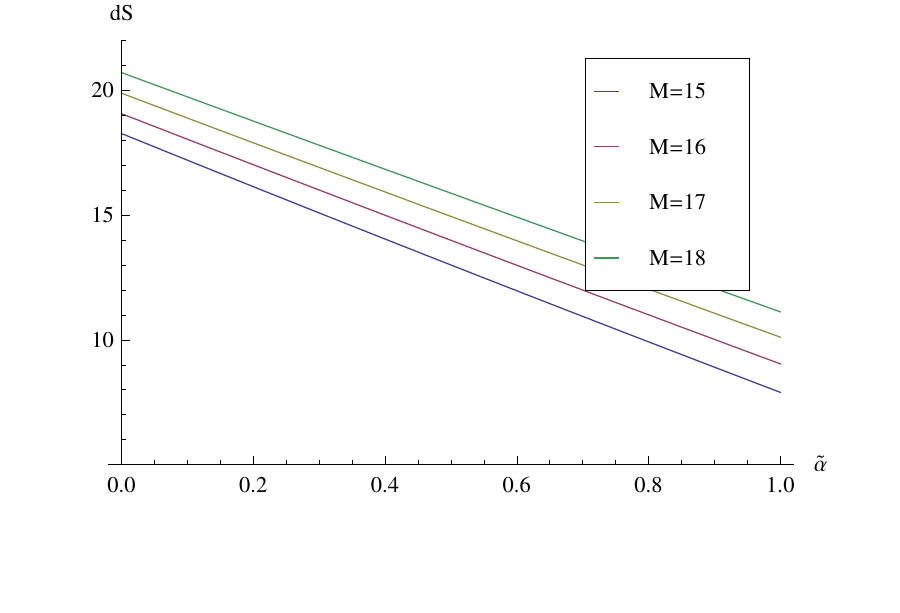}\\
  \caption{\scriptsize{The relation among $dS$, $M$ and $\widetilde{\alpha}$ for the case $k=1, \protect\\
  Q=5, \ell=1, d\ell=0.5 $.}}\label{f2}
 %\end{minipage}
 %\caption{²¢ÅÅͼÐÎ}
\end{minipage}%
\end{figure}

\begin{figure}[htp]
\begin{minipage}[t]{0.5\linewidth}
  \centering
  \setlength{\abovecaptionskip}{-0.1 cm}
  \setlength{\belowcaptionskip}{-0.4 cm}
  %\begin{minipage}[c]{0.5\textwidth}
  %\centering
  \includegraphics[width=0.65\textwidth,height=4.4cm]{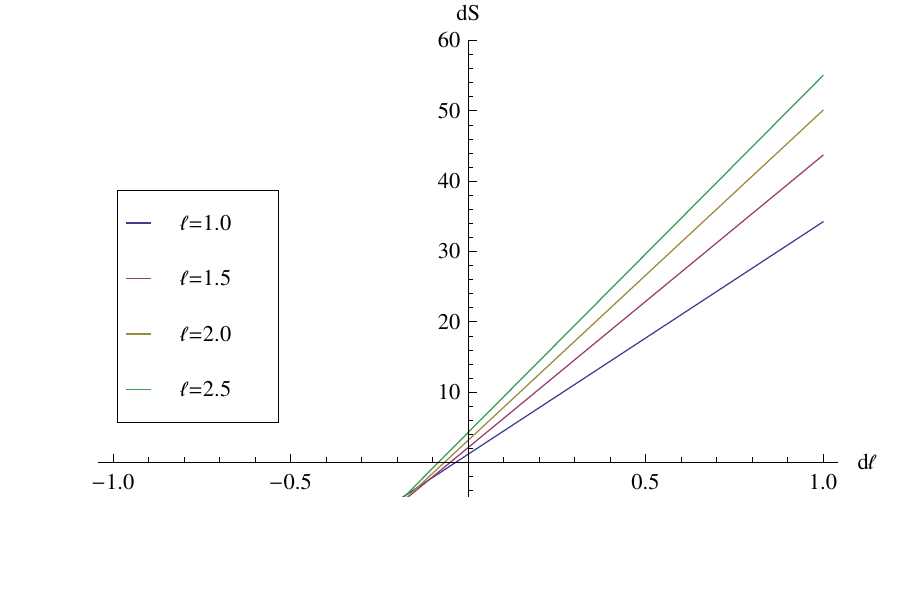}\\
  \caption{\scriptsize{The relation among $dS$, $\ell$ and $d\ell$ for the case $k=1, \protect\\
  \widetilde{\alpha} =0.5, Q =5, M =20$. In Fig.\ref{f3}, when $d\ell$ is smaller than some \protect\\
  certain value, $dS$ is a value less than zero.}}\label{f3}
 %\end{minipage}
 %\caption{²¢ÅÅͼÐÎ}
\end{minipage}%
\begin{minipage}[t]{0.5\linewidth}
  \centering
  \setlength{\abovecaptionskip}{-0.1 cm}
  \setlength{\belowcaptionskip}{-0.4 cm}
  %\begin{minipage}[c]{0.5\textwidth}
  %\centering
  \includegraphics[width=0.65\textwidth,height=4.4cm]{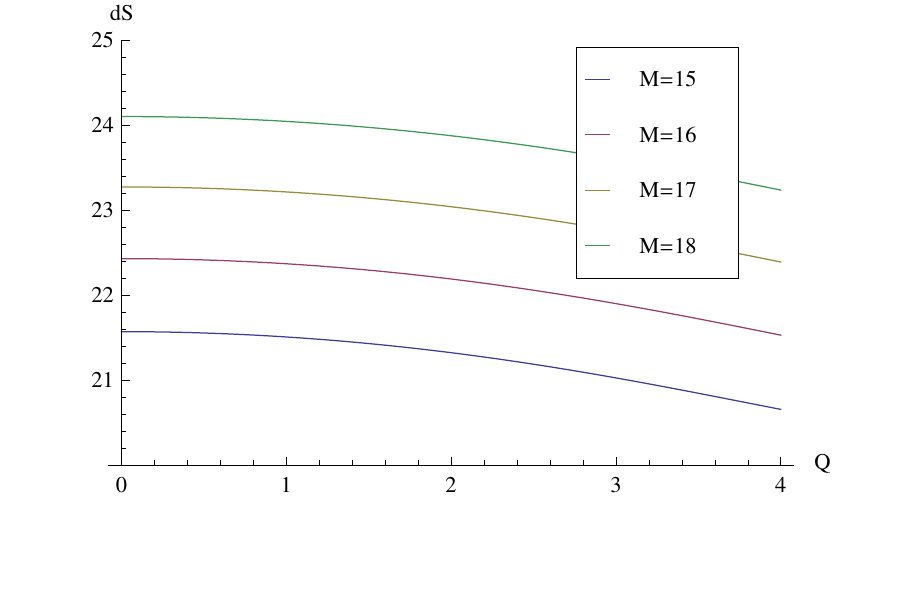}\\
  \caption{\scriptsize{The relation among $dS$, $M$ and $Q$ for the case $k=0, \protect\\
  \widetilde{\alpha} =0.5, \ell = 1, d\ell =0.5 $.}}\label{f4}
 %\end{minipage}
 %\caption{²¢ÅÅͼÐÎ}
\end{minipage}%
\end{figure}

\begin{figure}[htp]
\begin{minipage}[t]{0.5\linewidth}
  \centering
  \setlength{\abovecaptionskip}{-0.1 cm}
  \setlength{\belowcaptionskip}{-0.4 cm}
  %\begin{minipage}[c]{0.5\textwidth}
  %\centering
  \includegraphics[width=0.65\textwidth,height=4.4cm]{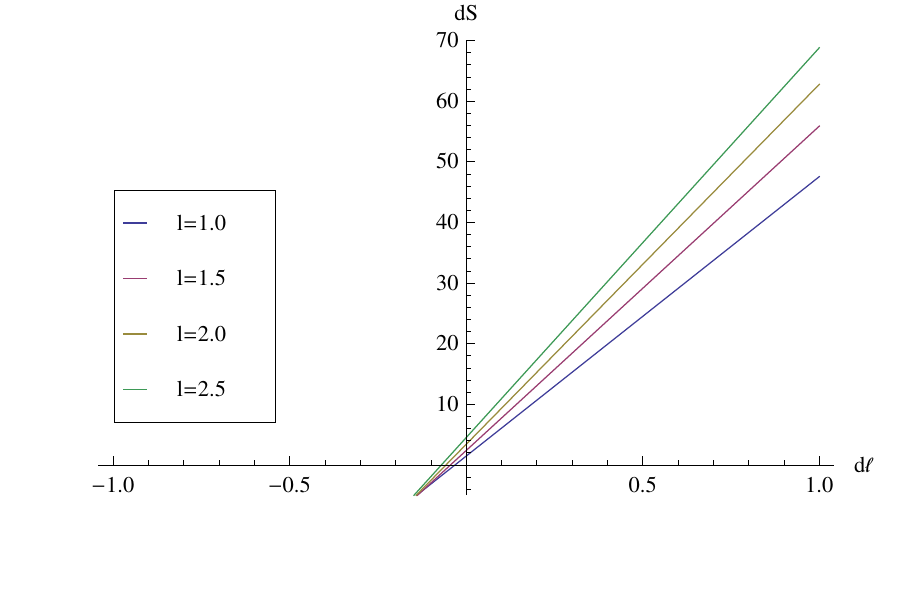}\\
  \caption{\scriptsize{The relation among $dS$, $\ell$ and $d\ell$ for the case $k=0,  \protect\\
  \widetilde{\alpha} =0.5, M = 20, Q =5$. In Fig.\ref{f5}, when $d\ell$ is smaller than some \protect\\
  certain value, $dS$ is a value less than zero.}}\label{f5}
 %\end{minipage}
 %\caption{²¢ÅÅͼÐÎ}
\end{minipage}%
\begin{minipage}[t]{0.5\linewidth}
  \centering
  \setlength{\abovecaptionskip}{-0.1 cm}
  \setlength{\belowcaptionskip}{-0.4 cm}
  %\begin{minipage}[c]{0.5\textwidth}
  %\centering
  \includegraphics[width=0.65\textwidth,height=4.4cm]{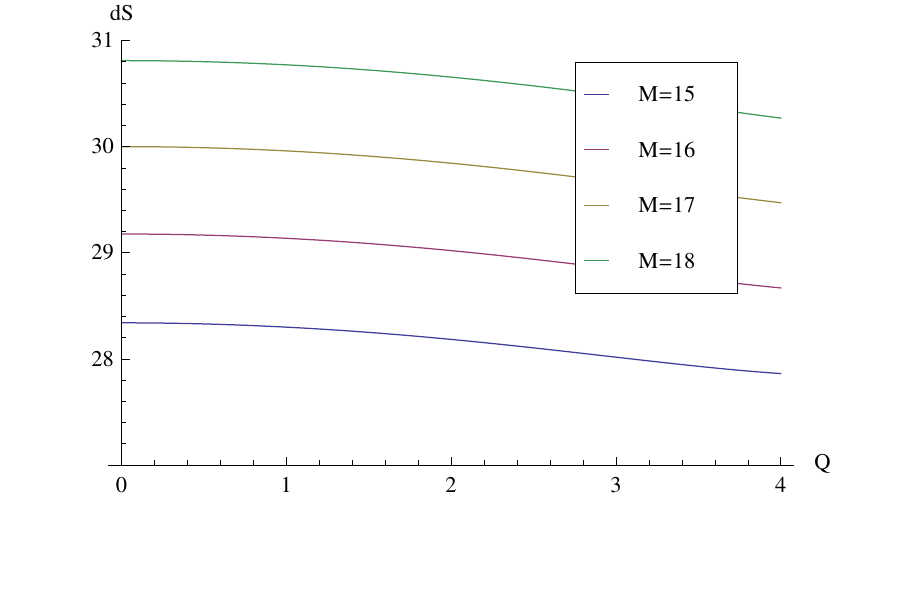}\\
  \caption{\scriptsize{The relation among $dS$, $M$ and $Q$ for the case $k=-1, \protect\\
  \widetilde{\alpha} =0.5, \ell = 1, d\ell =0.5 $.}}\label{f6}
 %\end{minipage}
 %\caption{²¢ÅÅͼÐÎ}
 \end{minipage}%
\end{figure}

\begin{figure}[htp]
\begin{minipage}[t]{0.5\linewidth}
  \centering
  \setlength{\abovecaptionskip}{-0.1 cm}
  \setlength{\belowcaptionskip}{-0.4 cm}
  %\begin{minipage}[c]{0.5\textwidth}
  %\centering
  \includegraphics[width=0.65\textwidth,height=4.4cm]{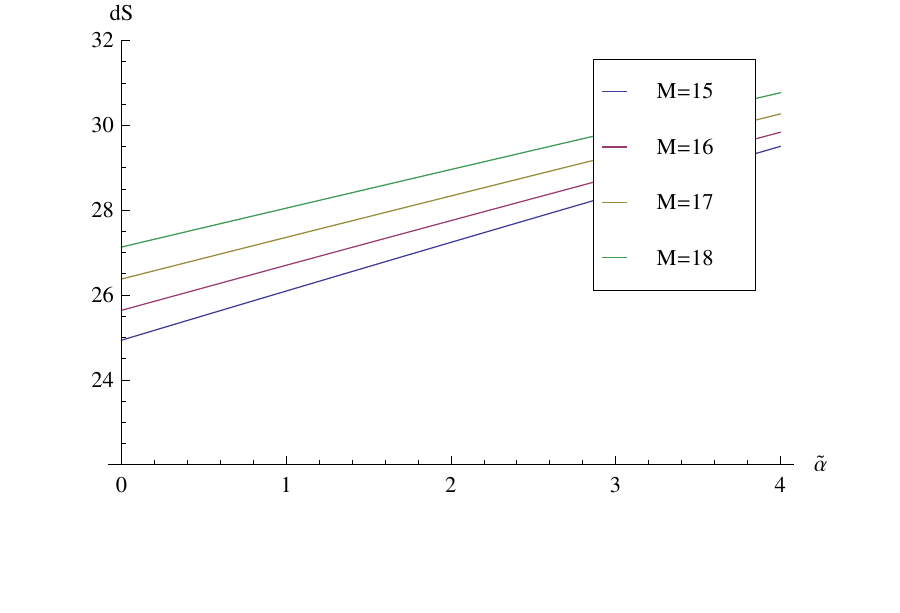}\\
  \caption{\scriptsize{The relation among $dS$, $M$ and $\widetilde{\alpha}$ for the case $k=-1, \protect\\
  Q=5, \ell=1, d\ell=0.5$.}}\label{f7}
 %\end{minipage}
 %\caption{²¢ÅÅͼÐÎ}
\end{minipage}%
\begin{minipage}[t]{0.5\linewidth}
  \centering
  \setlength{\abovecaptionskip}{-0.1 cm}
  \setlength{\belowcaptionskip}{-0.4 cm}
  %\begin{minipage}[c]{0.5\textwidth}
  %\centering
  \includegraphics[width=0.65\textwidth,height=4.4cm]{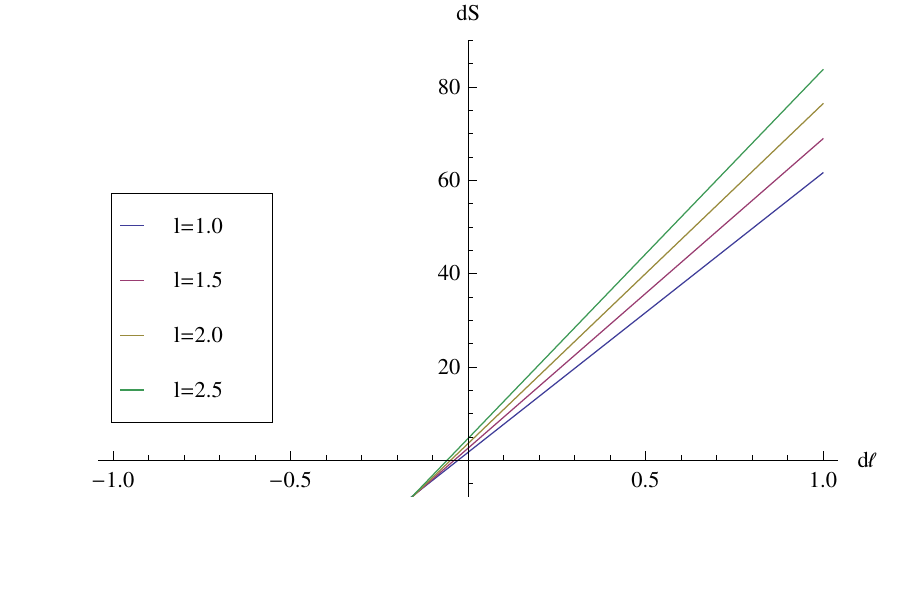}\\
  \caption{\scriptsize{The relation among $dS$, $\ell$ and $d\ell$ for the case $k=-1, \protect\\
  \widetilde{\alpha} =0.5, M = 20, Q =5$. In Fig.\ref{f8}, when $d\ell$ is smaller than some \protect\\
  certain value, $dS$ is a value less than zero.}}\label{f8}
 %\end{minipage}
 %\caption{²¢ÅÅͼÐÎ}
 \end{minipage}%
\end{figure}

In this section, the study of thermodynamics and weak cosmic censorship conjecture will be carefully addressed in the extended phase space. By viewing the cosmological constant as a dynamic variable, the first law of black hole thermodynamics in the extended phase space should be expressed as \cite{laws},
\begin{equation}\label{q27}
dM = TdS + \Phi dQ + VdP,
\end{equation}
where,
\begin{eqnarray}
P &=& -\frac{\Lambda}{8 \pi} = \frac{3}{8 \pi \ell^2}, \label{q28}
\\
V &=& \left( \frac{\partial M}{\partial P} \right)_{S,Q} = \frac{4 \pi r_+^3}{3},\label{q29}
\\
\Phi &=& \left( \frac{\partial M}{\partial Q} \right)_{S,P} =  \frac{Q}{r_+} \label{q30}
\end{eqnarray}
In this framework, we are going to check the first law of black hole thermodynamics when a charged particle dropped into black hole. Before that, it is worth noting that the mass $M$ is no longer the internal energy but the enthalpy of black hole, which is
\begin{equation}\label{q31}
M = U + PV,
\end{equation}
As stated in \cite{Page3}, the energy of an infalling particle should be the same amount as the mass(enthalpy) of black hole. So, we have
\begin{equation}\label{q32}
\omega = dM = dU + d(PV), q = dQ.
\end{equation}
With the help of Eq.(\ref{q6}), it leads to
\begin{equation}\label{q33}
dM = \Phi dQ + P^r,
\end{equation}
which is same as that described in the normal phase space. Similarly, the location of black hole horizon will be shifted when a particle dropped into the black hole. This shift $f(dr_+)$ can be expressed as
\begin{equation}\label{q34}
df_+ = \frac{\partial f_+}{\partial M} dM + \frac{\partial f_+}{\partial Q} dQ + \frac{\partial f_+}{\partial \ell} d\ell+ \frac{\partial f_+}{\partial r_+}dr_+ =0.
\end{equation}
Interestingly, we find that the terms $dM, dQ, d\ell$ are all disappear when one inserts Eq.(\ref{q33}) into Eq.(\ref{q34}). Finally, there is only the term $dr_+$ presented in Eq.(\ref{q34}), and it is
\begin{equation}\label{q36}
dr_+ = - \frac{ r_+^2 \left( r_+^3 d\ell + P^r \ell^3 \right) }{ \ell^3 \left( 2 Q^2 - 3 M r_+ + k ( r_+^2 - 4 k \widetilde{\alpha}) \right)}.
\end{equation}
Based on Eq.(\ref{q36}), the variations of the black hole entropy are
\begin{eqnarray}
dS &=&  - \frac{ 2\pi r_+ \left( r_+^3 d\ell + P^r \ell^3 \right) \left(  r_+^2 - 4 k \widetilde{\alpha} \right) }{ \ell^3 \left( 2 Q^2 - 3 M r_+ + k \left( r_+^2 - 4 k \widetilde{\alpha} \right) \right)}, \label{q37}
\end{eqnarray}
Using Eq.(\ref{q37}), it is easy to get
\begin{equation}\label{q38}
TdS - \frac{r_+^3 d\ell}{\ell^3}= P^r.
\end{equation}
Then, the equation (\ref{q33}) reduce to
\begin{equation}\label{q39}
dM = TdS + \Phi dQ + VdP.
\end{equation}
Obviously, the equation (\ref{q39}) is exactly the same as Eq.(\ref{q27}). This means that the first law of black hole thermodynamics still holds true in the extended phase space, which is coincide with that obtained in the normal phase space. Next, we will continue to check the second law of black hole thermodynamics in the extended phase space when a charged particle is captured by the black hole.
\\
\indent
As described in the normal phase space, it is clear that the temperature of the extremal black holes is equal to zero at the horizon. In this case, we have
\begin{equation}\label{q40}
dS_{e} \rightarrow\infty .
\end{equation}
Obviously, it is easy to see that the second law of black hole thermodynamics also holds true in the extended phase space, which is the same as that found in the normal phase space. For the non-extremal black holes, the variation of entropy $dS$ is presented as Eq.(\ref{q37}). In the extended phase space, the condition (\ref{q19}) still holds true. Considering this condition, and if the relationship ${d\ell} > -\left(P^r \ell^3\right)/{r_+^3}$ satisfied,
it is easy to find that the variation of entropy $dS$ is also positive, and thereby the second law of black hole thermodynamics does not violate in the extended phase space. This results is full in consistence with that obtained in the Reissner-Nordstr\"{o}m AdS black hole \cite{Page3}. To clearly see the value of $dS$, we have plotted Figs.\ref{f1}-\ref{f8} to show the relation between the variation of entropy $dS$ and other parameters\footnote{{Were, some accepted parameters have been employed to plot Figs.\ref{f1}-\ref{f8}, i.e., $\widetilde{\alpha} =0.5, \ell=1, p^r=1, Q=5, M = 20$. Obviously, as long as the condition (\ref{q19}) satisfied, other value of  $\widetilde{\alpha}, \ell, p^r, Q, M $ can also been used to present the variation of entropy. And, the reason of those parameters we employed in this paper is only to present the variation of entropy more clearly.}}.

In the extended phase space, for a black hole with a positive curvature horizon($k=1$), it is clear from Figs.\ref{f1},\ref{f2} that the variation of entropy $dS$ increases with the black hole mass $M$, but decreases with the black hole charge $Q$ and the the conformal anomaly parameter $\widetilde{\alpha}$. Similar to that in the normal phase space, one can easily check that there does not exist the effect of the conformal anomaly parameter $\widetilde{\alpha}$ on $dS$ for a black hole with a Ricci flat horizon($k=0$). And in this case, the variation of entropy is also bigger and bigger with the black hole mass $M$, but smaller and smaller with the black hole charge $Q$, which is shown in Fig.\ref{f4}. For a black hole with a negative constant curvature horizon($k=-1$), we find in Figs.\ref{f6},\ref{f7} that when the black hole mass $M$ and the parameter $\widetilde{\alpha}$ increase the value of $dS$ increases, but it decreases with the black hole charge $Q$. More importantly, it turns out in Figs.\ref{f3},\ref{f5},\ref{f8} that the variation of entropy $dS$ is always identified as a positive value as long as ${d\ell} > -\left(P^r \ell^3\right)/{r_+^3}$, otherwise is a value less than zero. That is to say, the second law of black hole thermodynamics
could be valid, and there is only a constraint on ${d\ell}$. In addition, one can see that the variation of entropy $dS$ always increases with the AdS radius $\ell$ and its variation $d\ell$, no matter what $k$ fixed.

Next, we will further check the weak cosmic censorship conjecture of the conformal anomaly corrected AdS black hole with particle's absorption. In a similar way, the mass, charge and the AdS radius of the black hole parameter will be changed into $( M+dM, Q+dQ, \ell + d\ell)$ when a particle drops into the black hole. And, the location of the event horizon, the minimum value and AdS radius will change into $( r_+ + dr_+, r_m+dr_m, \ell + d\ell)$. Correspondingly, the shift of the metric function $f(r)$ can be expressed as
\begin{eqnarray}\label{q42}
df_m &&= f(r_m + dr_m) - f_m \\
   &&= \frac{\partial f_m}{\partial M}dM + \frac{\partial f_m}{\partial Q}dQ + \frac{\partial f_m}{\partial \ell}d\ell,
\end{eqnarray}
where the condition $f'_m = 0$ has been used. For the extremal black hole, the location $r_m$ is exactly equal to the value of event horizon $r_+$. So, the relation (\ref{q33}) is still valid. After inserting Eq.(\ref{q33}) into Eq.(\ref{q42}), we have
\begin{equation}\label{q43}
df_m = - \frac{2 r_m}{r_m^2 - 4 k  \widetilde{\alpha}} \left( P^r + \frac{r_m^3d\ell}{\ell^3} \right).
\end{equation}
Using Eq.(\ref{q38}), it is easy to get
\begin{equation}\label{q44}
df_m = f_m + df_m = 0.
\end{equation}
It is obvious that Eq.(\ref{q44}) is same as Eq.(\ref{q24}), which means the weak cosmic censorship conjecture of the extremal black holes is still valid in the extended phase space. This results is coincide with that found in the normal phase space.

For the near-extremal black holes, the location $r_m$ is no longer equal to the event horizon $r_+$, which leads to that the condition (\ref{q33}) is not available. However, we note that $r_m$, $r_+$ are very close for the near-extremal black holes. So, in a similar way, the relation (\ref{q33}) can be expanded near the minimum point by using the relation $r_+ = r_m + \epsilon$, which is
\begin{eqnarray}\label{qq45}
dM = k dr_m - \frac{M dr_m}{r_m} + \frac{Q dQ}{r_m} - \frac{r_m^3 d\ell}{\ell^3} + \frac{2r_m^2 dr_m}{\ell^2} + \left( \frac{M dr_m}{r_m^2} - \frac{Q dQ}{r_m^2} - \frac{3 r_m^2 d\ell}{\ell^3} + \frac{4 r_m dr_m}{\ell^2}   \right) \epsilon +  O(\epsilon)^2.
\end{eqnarray}
Meanwhile, we note that the equation $f(r_+)=0$ can also be expanded with the relation $r_+ = r_m + \epsilon$. Then, it leads to
\begin{equation}\label{qq46}
Q = \frac{\sqrt{ 3r_m^4 + k r_m^2 \ell^2 + 2k^2 \ell^2 \widetilde{\alpha}}}{\ell},
\end{equation}
and
\begin{equation}\label{qq47}
dQ = \frac{- 3 r_m^4 d\ell + 6 \ell r_m^3 dr_m + k r_m \ell^3 dr_m }{\ell^2 \sqrt{3r_m^4 + kr_m^2 \ell^2 + 2k^2 \ell^2 \widetilde{\alpha}}}.
\end{equation}
With the aid of Eqs.(\ref{qq45}),(\ref{qq46}),(\ref{qq47}), it is easy to find the $df_m$ for the near-extremal black holes, which is
\begin{eqnarray}\label{q45}
%\begin{aligned}
&df_m =  O(\epsilon^2),\\
&f_m + df_m = \delta_\epsilon + O(\epsilon^2).
%\end{aligned}
\end{eqnarray}
From Eq.(\ref{q45}), it is true that there are always horizons for the near-extremal black holes. In this case, we can conclude that the configurations of the near-extremal black holes will not be changed in the extended phase space, thereby there has no violation of the weak cosmic censorship conjecture. Obviously, the result (\ref{q44}) for the extremal black holes can be naturally recovered by considering the condition($\epsilon \rightarrow 0$). Thus, it can be seen that the weak cosmic censorship conjecture are always valid for both the extremal black holes and near-extremal black holes.
This results are not only full in consistence with that found in \cite{Gwak,Chen,Zeng,Zeng1,Zeng2,Zeng3,Zeng4,Zeng5}, but also coincide with the findings obtained by using the new version of the gedanken experiments \cite{Many11,Many16,Many17,Many15}.
So in the context of the new assumption($\omega = dM$)\cite{Page3}, one can see that our results provide a piece of evidence for this new assumption \cite{Page3}.

\section{\textbf{Conclusions and Discussion}}\label{sec5}

In this paper, we have applied the particle's absorption method to carefully study the laws of thermodynamics and the weak cosmic censorship conjecture in the normal and extended phase space of conformal anomaly corrected AdS black hole. Firstly, we investigate the motion of a charged particle via the Hamilton-Jacobi equation, and obtain the relationship between the energy and momentum near the horizon. Then in the normal phase space, we have employed this relationship to further check the laws of black hole thermodynamics and the weak cosmic censorship conjecture as a charged particle dropped into black hole. It turns out that, the first law of black hole thermodynamics is well recovered from the equation (\ref{q17}). According to the variation of entropy $dS$, one can see that
the value of $dS$ is always positive for both the extremal black hole and the non-extremal black hole, which imply the second law of black hole thermodynamics is still hold by considering the particle's absorption. And by studying the shift of the matric function $df_m$, the results shows that, the value of $df_m$ is equal to zero for the extremal black hole, and $df_m$ always has a negative value for the near-extremal black hole. This means, there are always horizons to hidden the singularity, thereby the weak cosmic censorship conjecture is always valid, which is full in consistence with that obtained in \cite{Gwak,Chen,Zeng,Zeng1,Zeng2,Zeng3,Zeng4,Zeng5,Zeng6}.\\

Most of the resent papers claims that the second law of black hole thermodynamics violated by considering particle's absorption in the extended phase space. However, when assuming that the energy of an infalling particle changes should be the same amount as the mass(enthalpy) of black hole rather than only it's internal energy, we first find that the first law of black hole thermodynamics is well recovered by using the energy-momentum relationship (\ref{q6}), which is consistent with that obtained in \cite{Gwak,Chen,Zeng,Zeng1,Zeng2,Zeng3,Zeng4,Zeng5,Zeng6}. Then in this case, it further turns out that, the variation of entropy $dS$ are also positive for both the extremal black hole and the non-extremal black hole when the condition ${d\ell} > -\left(P^r \ell^3\right)/{r_+^3}$ satisfied. So, the second law that entropy always increases could hold, which means it can also be the supreme position among the laws of Nature. Contrary to the previous studies, this conclusion is a more physical result in the extended phase space. Finally in the extended phase space, our further discussion on the weak cosmic censorship conjecture shows that the black hole horizons always exists and the singularity is also covered by it. So in the context of the new assumption, there is no violation of the weak cosmic censorship conjecture for the conformal anomaly corrected AdS black hole, which is full in consistence with that obtained in \cite{Many11,Many16,Many17,Many15,Gwak,Chen,Zeng,Zeng1,Zeng2,Zeng3,Zeng4,Zeng5}. Therefore, it can be seen that this result provide a piece of evidence of the fact that the energy of an infalling particle changes indeed is the same amount as the mass(enthalpy) of black hole.
In addition, we note that the conformal anomaly parameter $\widetilde{\alpha}$ and the the scalar curvature parameter $k$ do not have any influence on our conclusions with respect to the laws of thermodynamics and the weak cosmic censorship conjecture.

\setlength{\parindent}{0pt}\textbf{\textbf{Acknowledgments}}
{The authors would like to thank the anonymous reviewers for their helpful comments and suggestions, which helped to improve the quality of this paper.} This work is supported by the National Natural Science Foundation of China (Grant No.11903025), and by the starting fund of China West Normal University (Grant No.18Q062), and by the Natural Science Foundation of Sichuan Education Department (Grant No.17ZA0294), and by the Research Project of Si Chuan MinZu College (Grant No.XYZB18003ZA).\\

%
% For  figures use
%\begin{figure*}
% Use the relevant command for your figure-insertion program
% to insert the figure file. See example above.
% If not, use
%\vspace*{5cm}       % Give the correct figure height in cm
%\includegraphics{leer.eps}
%\caption{Please write your figure caption here}
%\label{fig:2}       % Give a unique label
%\end{figure*}
% or  this
%\begin{figure}
%\centering
% Use the relevant command for your figure-insertion program
% to insert the figure file.
% For example, with the option graphics use
%\resizebox{0.75\textwidth}{!}{%
%  \includegraphics{leer.eps}
%}
% If not, use
%\vspace{5cm}       % Give the correct figure height in cm
%\caption{Please write your figure caption here}
%\label{fig:1}       % Give a unique label
%\end{figure}
%
%
% For tables use
%\begin{table}
%\centering
%\caption{Please write your table caption here}
%\label{tab:1}       % Give a unique label
% For LaTeX tables use
%\begin{tabular}{lll}
%\hline\noalign{\smallskip}
%first & second & third  \\
%\noalign{\smallskip}\hline\noalign{\smallskip}
%number & number & number \\
%number & number & number \\
%\noalign{\smallskip}\hline
%\end{tabular}
% Or use
%\vspace*{5cm}  % with the correct table height
%\end{table}

%
% BibTeX users please use
% \bibliographystyle{}
% \bibliography{}
%
% Non-BibTeX users please use

\end{document}